\documentclass[12pt]{article}
\usepackage{amsmath} 
\usepackage{graphicx}
\usepackage{hyperref}
\usepackage{bbold}
\usepackage{lipsum}
\usepackage{braket}
\usepackage{cases}
\usepackage{dsfont,amsmath,amssymb,color,units,pstricks}
\usepackage{amsfonts}
\usepackage{epsfig}
\usepackage{bm}
 \usepackage{graphicx}
\usepackage[utf8]{inputenc} 
\usepackage[T1]{fontenc}      
\usepackage{textcomp}            
\usepackage{enumitem}
\newcommand \be  {\begin{equation}}
\newcommand \bea {\begin{eqnarray} \nonumber }
\newcommand \ee  {\end{equation}}
\newcommand \eea {\end{eqnarray}}

\usepackage[bitstream-charter, greekfamily=default]{mathdesign}

\title{Excess Out-of-Sample Risk and Fleeting Modes}
\author{Jean-Philippe Bouchaud, Iacopo Mastromatteo,\\ Marc Potters \& Konstantin Tikhonov \\ {\small{
Capital Fund Management, 23 rue de l'Universit\'e,}}\\ {\small{75007 Paris, France}}}

\date{April 2022}

\begin{document}

\maketitle
\begin{abstract}
    Using Random Matrix Theory, we propose a universal and versatile tool to reveal the existence of ``fleeting modes'', i.e. portfolios that carry statistically significant excess risk, signalling ex-post a change in the correlation structure in the underlying asset space. Our proposed test is furthermore independent of the ``true'' (but unknown) underlying correlation structure. We show empirically that such fleeting modes exist both in futures markets and in equity markets. We proposed a metric to quantify the alignment between known factors and fleeting modes and identify momentum as a source of excess risk in the equity space. 
\end{abstract}

\paragraph{Introduction} Managing the risk of large portfolios requires the knowledge of equally large covariance matrices, describing the whole array of pairwise cross-correlation between the assets included in the portfolio. 

As is well-known by now, the empirical determination of such covariance matrices is difficult -- for at least two different reasons. One is that even in a stationary world, that is, a world described by an unknown underlying stochastic process with time independent parameters, empirical covariance matrices are soiled by a large amount of measurement noise, that only goes to zero as $\sqrt{N/T}$, where $N$ is the number of assets and $T$ the amount of data points in the time direction -- for a recent review, see \cite{book}. Typical numbers are $N=500$ stocks in a portfolio, and $T=1000$ days of data (corresponding to $4$ years), $\sqrt{N/T} \approx 0.7$ which is by no means small! A number of techniques have been proposed over the years to ``clean'' as efficiently as possible the empirical covariance matrix $\mathbb{E}$ such that one approaches as well as possible the ``true'' covariance matrix $\mathbb{C}$ \cite{LedoitWolf1,ElKaroui,LedoitPeche}, and for reviews \cite{Bun,LedoitWolf2} and refs.\ therein. Such cleaning schemes, some based on sophisticated Random Matrix Theory techniques, do help in reducing the discrepancy between ``out-of-sample'' risk (i.e. risk realized in a period outside the training sample) and ``in-sample'' risk (i.e. risk estimated on the same period as the training sample). 

However, the assumption of a stationary world is certainly too naive to describe financial markets. For one thing, volatility can strongly fluctuate from one period to the next, so ``out-of-sample'' risk may be larger or smaller than ``in-sample'' risk simply due to realized volatility. This is a well-studied issue, which can be partly mitigated by the use of sophisticated volatility models and/or using the forward looking, implied volatility from option markets. In this study, we are rather concerned about  {\it correlation risk}. As a striking example, think of the correlation between the daily price changes of the S\&P500 index and the US T-Bond. For many years before 1997, it hovered around $+0.5$, before suddenly switching sign around the so-called Asian crisis. It then remained in negative territory -- in a ``flight-to-safety'' mode -- for more than 20 years before possibly switching again in 2021/2022, time will say \cite{seager}. More generally, one can expect that as macroeconomic conditions evolve, the whole correlation structure between financial assets also evolves. Several ideas to quantify such a genuine evolution of correlations have been discussed in the past \cite{BEKK,Reigneron,Allez,Karami}, in particular the interesting notion of ``market states'' \cite{MarketStates1,MarketStates2}. 

The main difficulty is to disentangle measurement noise, which leads to an apparent evolution of the empirical covariance matrix between two non overlapping periods, from any possible evolution of the underlying covariance matrix $\mathbb{C}$. In Ref. \cite{overlaps}, two of us proposed a non parametric method based on the overlap of the eigenvectors of $\mathbb{E}_{\text{in}}$ and those of $\mathbb{E}_{\text{out}}$, where ``in'' and ``out'' refer, respectively, to the in-sample and out-of-sample period. Quite interestingly, our proposal did not require the knowledge of $\mathbb{C}$, only that it was time independent.   
In this note, we want to propose an alternative non parametric test, simpler and more transparent, which again does not rely on the knowledge of $\mathbb{C}$ and allows one to diagnose periods of statistically significant excess correlation risk and identify the directions (in asset space) along which such excess risk manifests itself.

\paragraph{Theoretical Tools \& Analytical Results} Let $\mathbf{X} = X_{i,t}$ be the return data set, where $i$ is the asset label and $t$ the time label. 
The in-sample covariance matrix $\mathbb{E}_{\text{in}}$ is defined as 
\be
\label{eq:averaging}
\mathbb{E}_{\text{in}} = \frac{1}{T_{\text{in}}} \sum_{t \in \text{in}} X_{i,t} X_{j,t} :=  \frac{1}{T_{\text{in}}} \mathbf{X}_{\text{in}} \mathbf{X}_{\text{in}}^\top, 
\ee 
where $T_{\text{in}}$ is the length of the in-sample period. The out-of-sample covariance matrix $\mathbb{E}_{\text{out}}$ is defined similarly, with $T_{\text{out}}$ is the length of the out-sample period.

Let us now introduce the matrix $\mathbb{D}$ defined as:
\be
\mathbb{D} := \mathbb{E}_{\text{in}}^{-1/2} \mathbb{E}_{\text{out}} \mathbb{E}_{\text{in}}^{-1/2},
\ee
where $\mathbb{E}_{\text{in}}^{1/2}$ is defined as 
the {\it symmetric} matrix square-root of $\mathbb{E}_{\text{in}}$. The intuitive meaning of $\mathbb{D}$ is as follows. By defining $\mathbf{Y}_{\text{in}} := \mathbb{O} \mathbb{E}_{\text{in}}^{-1/2} \mathbf{X}_{\text{in}}$, where $\mathbb{O}$ is an arbitrary rotation matrix, we construct a set of $N$ synthetic assets (or portfolios) that are by construction ortho-normal, i.e. each synthetic asset is of unit risk and uncorrelated (in sample) with all other synthetic assets. 

Now the \textit{out-of-sample} covariance matrix of these synthetic assets is given by $\mathbb{D}_\mathbb{O}:=\mathbb{O} \mathbb{D} \mathbb{O}^\top$, whose  eigenvectors specify a new set of {\it uncorrelated} synthetic assets, with variance given by the eigenvalues $\lambda_a$ (which are independent of $\mathbb{O}$). Since the in-sample risk of the synthetic assets has been normalized to one, the eigenvectors associated with the eigenvalues $\lambda_a > 1$ thus correspond to linear combinations of synthetic assets that over-realize their risk in the out-of-sample period.  
In the following, we will choose $\mathbb{O}$ such that the synthetic assets are simply the principal risk components $\vec{v}_\mu$ of the in-sample covariance matrix $\mathbb{E}_{\text{in}}$. We will call the directions $\vec{v}_\mu$ the {\it statistical risk modes}. 

Suppose $q_{\text{in}}=N/T_{\text{in}}$ and $q_{\text{out}}=N/T_{\text{out}}$ are both very small and the world is stationary, with true covariance matrix $\mathbb{C}$. Then, clearly
\be 
\mathbb{E}_{\text{in}} \approx \mathbb{E}_{\text{out}} \approx \mathbb{C} \longrightarrow
\mathbb{D} \approx \mathbb{I}.
\ee
Hence, in this case, all eigenvalues of $\mathbb{D}$ are very close to unity -- no portfolio over-realizes its risk, as expected. In the case where $q_{\text{in}}$ and $q_{\text{out}}$ take arbitrary values, one first notes that by definition, (white) Wishart matrices $\mathbb{W}$ correspond to empirical covariance matrices when $\mathbb{C}=\mathbb{I}$. Hence one can write, still assuming stationarity:
\be
\mathbb{E}_{\text{in}} = \mathbb{C}^{1/2} \mathbb{W}_{\text{in}} \mathbb{C}^{1/2}; \qquad
\mathbb{E}_{\text{out}} = \mathbb{C}^{1/2} \mathbb{W}_{\text{out}} \mathbb{C}^{1/2}.
\ee 
Now, since the characteristic polynomial of $\mathbb{D}$ is the same as that of $\mathbb{E}_{\text{in}}^{-1} \mathbb{E}_{\text{out}} =
\mathbb{C}^{-1/2} \mathbb{W}_{\text{in}}^{-1} 
\mathbb{W}_{\text{out}} \mathbb{C}^{1/2}$, which is turn is the same as that of $\mathbb{W}_{\text{in}}^{-1} 
\mathbb{W}_{\text{out}}$, we conclude that the eigenvalues of $\mathbb{D}$ are actually {\it independent} of $\mathbb{C}$, and equal to those of our theoretical benchmark 
\be 
\mathbb{D}_{\text{th.}} := \mathbb{W}_{\text{in}}^{-1/2} 
\mathbb{W}_{\text{out}} \mathbb{W}_{\text{in}}^{-1/2},
\ee 
where $\mathbb{W}_{\text{in}}$, $\mathbb{W}_{\text{out}}$ are independent Wishart matrices of parameter, respectively, $q_{\text{in}}$ and $q_{\text{out}}$. In the following, we will assume $q_{\text{in}}<1$ , i.e. $T_{\text{in}} > N$, so that  
$\mathbb{W}_{\text{in}}$ is invertible.

The matrix $\mathbb{D}_{\text{th.}}$, a close relative of Jacobi random matrices, is the product of a Wishart and inverse-Wishart matrix and its spectrum can easily be computed, see e.g. \cite{book}.  Denoting $\lambda$ its eigenvalues, the probability density function of $\lambda$ reads:
\be \label{rho_lambda}
\rho(\lambda) = \frac{1-q_{\text{in}}}{2\pi}\frac{\sqrt{[(\lambda_{\max} - \lambda)(\lambda - \lambda_{\min})]^+} }{\lambda (q_{\text{in}}\lambda+q_{\text{out}})}+\left[1-q_{\text{out}}^{-1}\right]^+\delta(\lambda)
\ee 
with
\be
\lambda_{\max,\min} = \frac{1+q_{\text{in}}+q_{\text{out}}(1-q_{\text{in}})
\pm 2 \sqrt{q_{\text{in}}+q_{\text{out}}(1-q_{\text{in}})}}{(1-q_{\text{in}})^2},
\ee
where the symbol $[]^+$ denotes the positive part. Note that for $q_{\text{out}}>1$, a finite fraction of eigenvalues are exactly zero as expressed by the Dirac delta function.
This density has mean $(1-q_{\text{in}})^{-1}$ and variance $(1-q_{\text{in}})^{-3}(q_{\text{in}}+q_{\text{out}}(1-q_{\text{in}}))$. 

\begin{figure}
    \centering
    \includegraphics{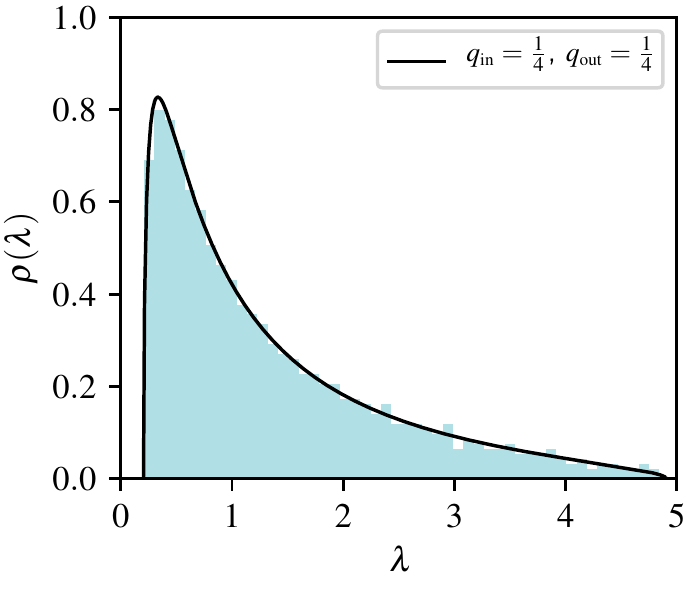}
    \hspace{1cm}
    \includegraphics{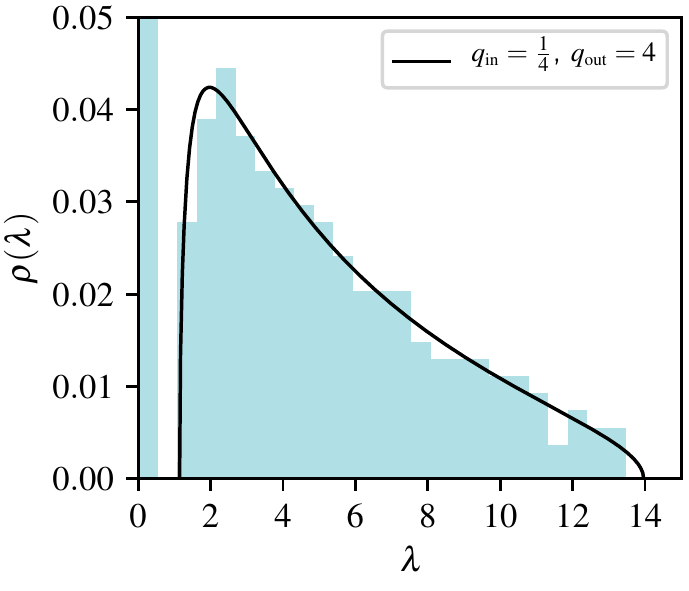}
    \caption{Density of eigenvalue of the matrix $\mathbb{D}_{\text{th.}}$ (Eq.\ \eqref{rho_lambda}) for $q_{\text{in}}=1/4$ and $q_{\text{out}}=1/4$ (left) and $q_{\text{out}}=4$ (right) compared with a numerical simulation with $N=1000$. Note that when $q_{\text{out}}>1$ as on the right, there is a Dirac at zero with weight $1-q_{\text{out}}^{-1}$. Note also that the numerical histogram drops to zero slightly below the theoretical value $\lambda_{\max}$. This is a finite $N$ effect, see also Fig. \ref{fig:avg_spectrum}.}
    \label{fig:rho_simu}
\end{figure}

\begin{figure}
    \centering
    \includegraphics[width=\textwidth]{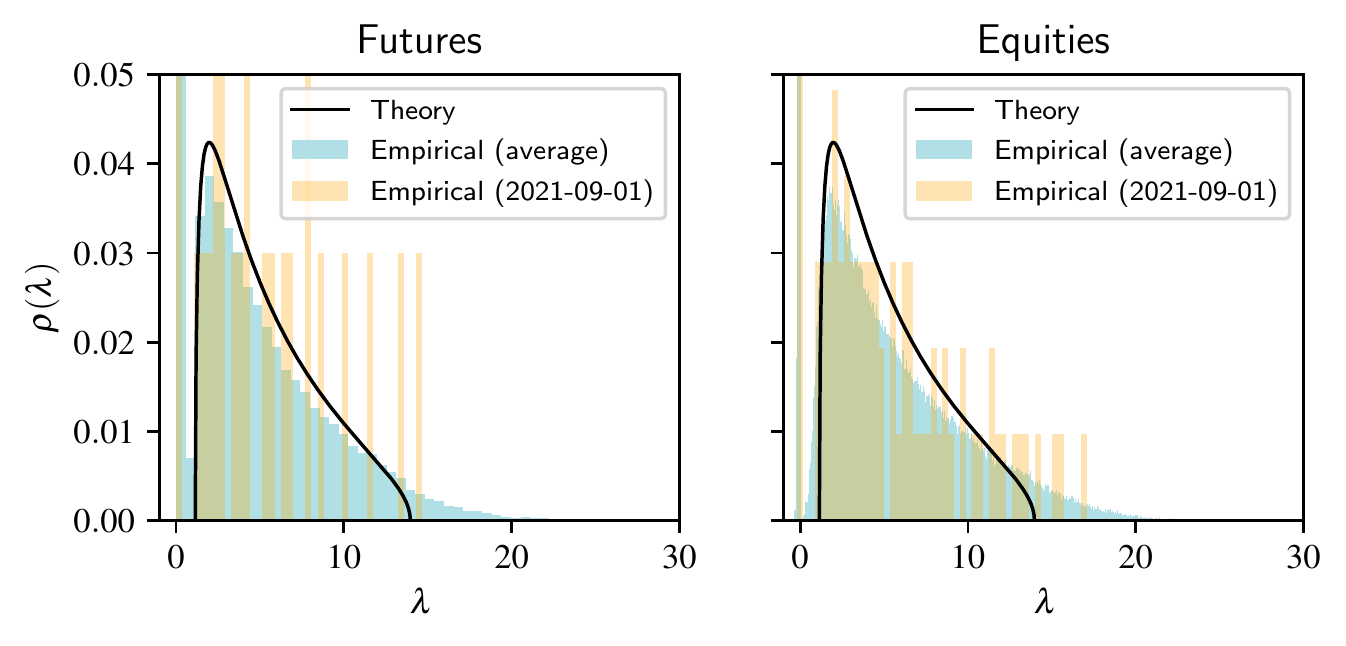}
    \caption{
	Average eigenvalue density for the matrix $\mathbb{D}(t)$ on futures (left panel) and stocks (right panel).
	In both cases the spectrum predicted by Eq.~\eqref{rho_lambda} for $q_{\text{in}}=1/4$ and $q_{\text{out}}=4$ is plotted for reference against the average empirical density. Eq. \eqref{eq_lambdamax} yields $\lambda_{\min}= 1.15$ and $\lambda_{\max}= 13.97$ but finite $N$ effects are expected to shift downwards and blur the right edge $\lambda_{\max}$ on a scale $\Delta_N = (c N)^{-2/3}$ (see e.g. \cite{book}, ch. 14), with $c \approx 2.7 \, 10^{-3}$, leading to $\Delta_N \approx 2.4$ for futures ($N=98$) and $\Delta_N \approx 1.15$ for stocks ($N=300$).
	The spectrum of a randomly chosen day (2021-09-01) is also shown for comparison for the two data sets. 
	}
	    \label{fig:avg_spectrum}
\end{figure}

Our null hypothesis test is thus the following: if the true underlying covariance matrix $\mathbb{C}$ is the same in-sample and out-of-sample, the non-zero eigenvalues $\lambda$ of $\mathbb{D}$ should, for large $N$, all lie within the interval\footnote{For finite $N$, there are corrections to $\lambda_{\max,\min}$ of order $N^{-2/3}$, with a prefactor that can be large in practice, see Fig. \ref{fig:avg_spectrum}.}
\be 
\lambda \in \left[\lambda_{\min},\lambda_{\max}\right],
\ee 
with a distribution compatible with Eq. \eqref{rho_lambda}, see Fig. \ref{fig:rho_simu} for a particular illustration and numerical simulations.
Several limiting cases are interesting to discuss. One is when the in-sample and out-of-sample period have the same length, i.e. $T_{\text{in}}=T_{\text{out}}=T$, or $q_{\text{in}}=q_{\text{out}}=q$. One then finds
\be \label{eq_lambdamax}
\lambda_{\max,\min} = \frac{1+2q-q^2\pm 2 \sqrt{q(2-q)}}{(1-q)^2}
\ee 
When both periods are very long compared to $N$, one has $q \to 0$ and therefore the interval where the eigenvalues of $\mathbb{D}$ are expected to be found is
\be 
\lambda \in \left[
1 - 2\sqrt{2q}, 
1 + 2\sqrt{2q}\right], \qquad q := \frac{N}{T},
\ee 
which, as expected, tends to a Dirac mass at $\lambda=1$ for $q=0$.

Now, look at another interesting regime where $T_{\text{in}} \gg T_{\text{out}} > N$, i.e. long in-sample period and relatively short out-of-sample period, aiming at detecting abrupt ``regime shifts''. In this regime where $q_{\text{in}} \to 0$, we recover precisely the Mar\v{c}enko-Pastur distribution with parameter $q=q_{\text{out}}$ \cite{book}, as it should be since in that limit $\mathbb{W}_{\text{in}} \equiv \mathbb{I}$ and $\mathbb{E}_{\text{in}} \equiv \mathbb{C}$. The non-zero eigenvalues of $\mathbb{D}_{\text{th.}}$ satisfy,
\be 
\lambda \in \left[(1-\sqrt{q_{\text{out}}})^2,(1+\sqrt{q_{\text{out}}})^2\right],
\ee 
with an additional Dirac mass of weight 
$1-q_{\text{out}}^{-1}$ when $q_{\text{out}} > 1$.

\paragraph{Empirical Analysis for Stocks \& Futures} In order to quantify by how much real financial returns differ from their stylized counterpart above, we have constructed two data sets consisting of
daily returns $\mathbf{X} = X_{i,t}$ of two different groups of financial instruments. The first data set comprises a set of $N=98$ liquid futures, covering
different sectors (stock indices, commodities, FX, yields), expiry dates and geographies (America, Europe, Asia, and a smaller set of developing markets)
in a period ranging from 2006-01-01 to 2022-03-01.
The second data set consists in $N=300$ US stocks in the period 2002-05-27 to 2022-01-21.\footnote{The detailed list of futures contracts and US stocks is available upon request to the authors. Note that we have {\it not} detrended daily returns by their means, which are for all purposes here negligible.} In order to get rid of any spurious volatility fluctuations and only focus on correlations, we normalize each daily returns by its own intraday volatility, constructed using a Garman-Klass estimator based on Close-High-Low-Close data. 

In both universes, we construct a set of rolling estimators $\mathbb{E}_{\text{in}}(t)$ and $\mathbb{E}_{\text{out}}(t)$ according to the following prescription.
First, for each day after a ``burning'' period $t > T_{\text{in}} + T_{\text{out}}$, we consider the $(\text{in})$ interval as the one comprising returns belonging to $[ t - T_{\text{out}} - T_{\text{in}}, t - T_{\text{out}}[$,
whereas the $(\textrm{out})$ interval is built with the returns belonging to $[t-T_{\text{out}}, t[$. This construction ensures that i) intervals built at time $t$
only use data available at day $t$ ii) the intervals are contiguous but perfectly disjoint iii) all in-sample and out-of-sample intervals have exactly the same lengths $T_{\text{in}}$ and $T_{\text{out}}$
iv) under the hypothesis of i.i.d.\ returns, the estimators $\mathbb{E}_{\text{in}}(t)$
and $\mathbb{E}_{\text{out}}(t)$ are distributed according to the null model described above.

For both futures and stocks, we have decided to fix $q_{\textrm{in}} = \frac 1 4$ and $q_{\text{out}} = 4$, which corresponds to an in-sample period of a year and a half
for futures (about five years for stocks) and an out-of-sample interval of approximately one month for futures (slightly less than four months for stocks). We then apply the definition  Eq.~\ref{eq:averaging} to both in-sample and out-of-sample intervals, obtaining rolling sets of covariance matrices  indexed by $t$ and denoted as $\mathbb{E}_{\text{in}}(t)$ and $\mathbb{E}_{\text{out}}(t)$.

Finally, we build risk over-realization matrices $\mathbb{D}(t) = \mathbb{E}_{\text{in}}^{-1/2}(t) \, \mathbb{E}_{\text{out}}(t) \, \mathbb{E}_{\text{in}}^{-1/2}(t)$,
from which we can extract eigenvalues $\left\{ \lambda_{a}(t) \right\}$ that can be compared to the ones prescribed by our null hypothesis. The corresponding eigenvectors also contain important information, to be discussed below. 

\begin{figure}
    \centering
    \includegraphics[width=\textwidth]{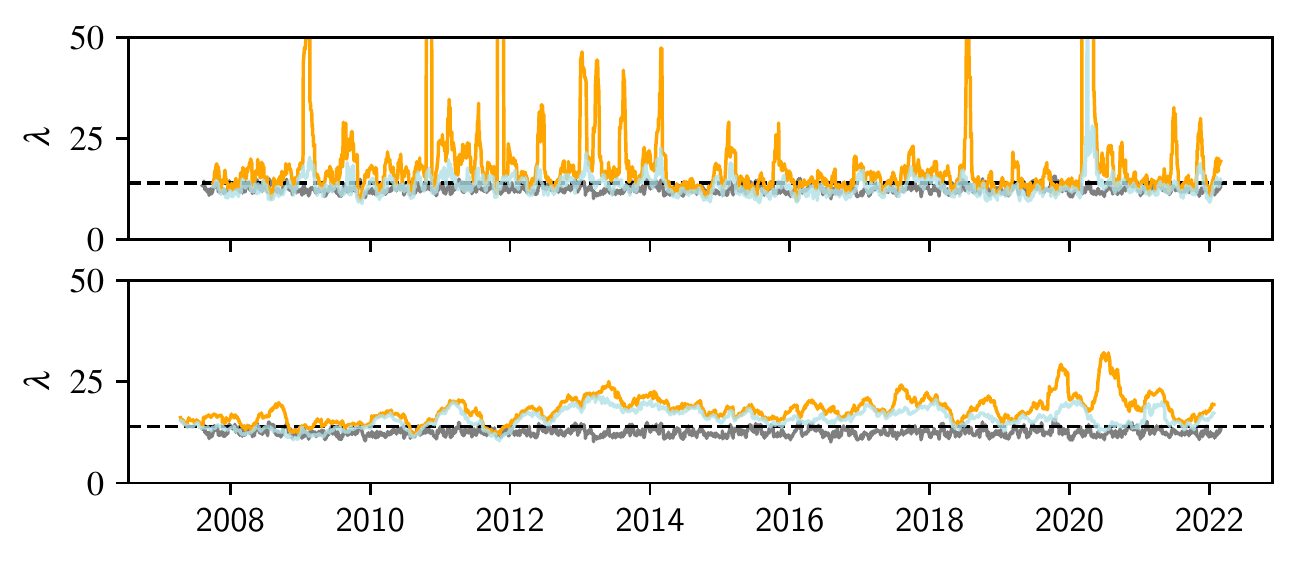}
    \caption{
	Time trajectory of the largest two eigenvalues $\lambda_{1}(t)$ (orange line) and $\lambda_{2}(t)$ (light blue line) for futures (top panel) and stocks (bottom panel),
	indicating that departures from the null model for the top eigenvalue $\lambda_{\max} = 13.97$ (dashed line, see Eq. \eqref{eq_lambdamax}) are significant for a substantial fraction of days. Note that the simulated null model with Gaussian returns (grey line) slightly undershoots the theoretical value of $\lambda_{\max}$ because of finite $N$ effects.
	}
	    \label{fig:lambda_vs_t}
\end{figure}

Fig.~\ref{fig:avg_spectrum} illustrates the result of such comparison for both futures (left panel) and stocks (right panel), indicating that in both cases we detect significant departures from the null model. However, the average eigenvalue distribution does not distinguish between an intermittent scenario where risk-over realisation is clustered in time, from a uniform scenario where risk is always over-realized. Part of such information is summarized in Fig.~\ref{fig:lambda_vs_t}, where the evolution of the top eigenvalue
$\lambda_1(t)$ is displayed for both our data sets, and compared to the value expected under our null model. Departure from the null model are relatively mild in some periods and stronger in others, with clear spikes, notably in the futures space.

The directions along which excess out-of-sample risk is large are given by the eigenvectors $\vec{z}_a$ of $\mathbb{D}_{\mathbb{O}} := \mathbb{Z}^\top \textrm{diag}({\lambda}_a) \mathbb{Z}$ corresponding to the largest eigenvalues $\lambda_a$. As explained above, because of our choice of $\mathbb{O} : = \vec{v}$, these eigenvectors can be interpreted as portfolios of the (in-sample) statistical risk modes $\vec{v}_\mu$, $\mu=1, \dots, N$. We will call the eigenvectors $\vec{z}_a$ {\it fleeting modes}.

\begin{figure}
    \centering
    \includegraphics[width=\textwidth]{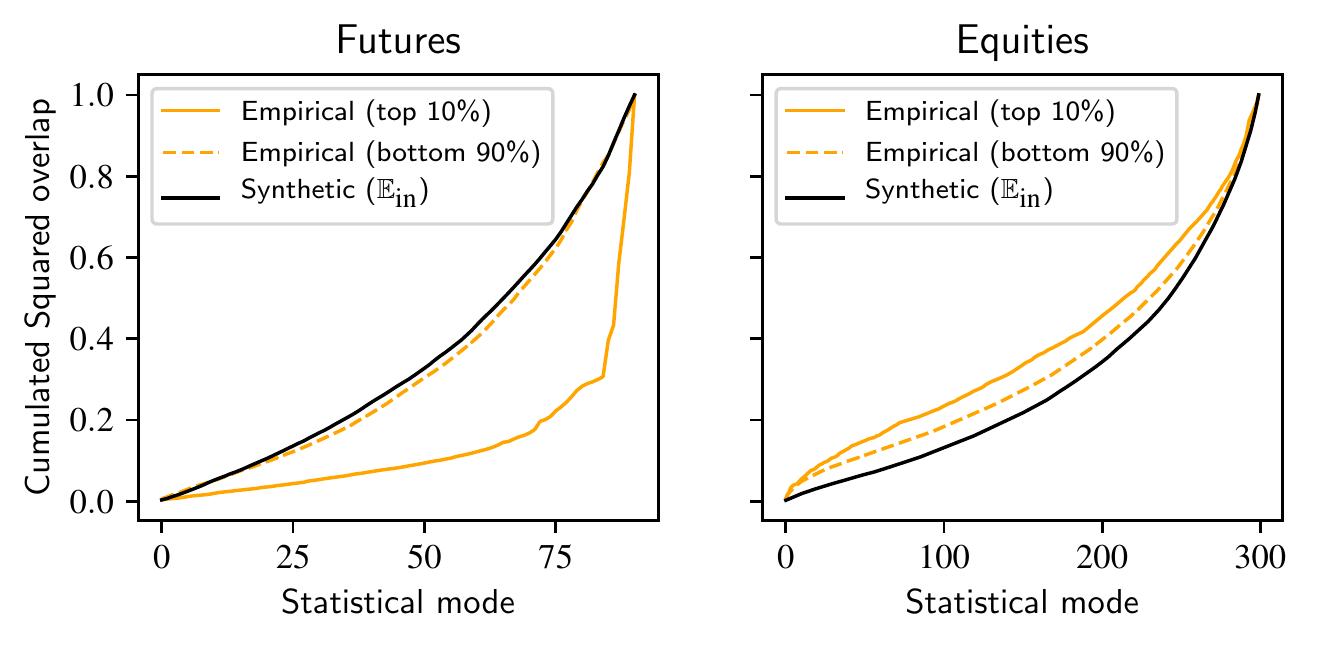}
    \caption{
    Average of the cumulative squared overlap between the first fleeting mode $\vec z_1$ and the statistical risk modes $\vec v_\mu$, for the top $10 \%$ cases of risk over-realisation (plain orange lines) and the bottom $10 \%$ cases of risk over-realisation (dashed lines). The black line corresponds to a stationary null hypothesis where the true covariance matrix is equal to the in-sample covariance matrix $\mathbb{E}_{\text{in}}$. (Left) futures; (Right) Equities. The $x$-axis is the rank $n$ of the corresponding statistical mode, low rank meaning large in sample risk. Note that risk over-realisation is concentrated in directions that are markedly different in the two cases.   
	}
	    \label{fig:overlaps}
\end{figure}

The first question one would like to ask is how close is the top fleeting mode $\vec{z}_1$ (corresponding to the top eigenvalue $\lambda_1$) to the dominant in-sample risk modes $\vec{v}_\mu$. 
We thus define the cumulative squared overlap as $\psi_n:=\sum_{\mu=1}^n (\vec {z}_{1} \cdot \vec \nu_\mu)^2 $, where $\vec {z}_{1} \cdot \vec \nu_\mu$ is the $\mu$th risk mode component of $\vec {z}_{1}$. Note that $\psi_N \equiv 1$ for $n=N$, because the set of $\{ \vec{v}_\mu \}$ forms an ortho-normal basis. Fig. \ref{fig:overlaps} shows $\psi_n$ as a function of $n$, both for futures and for equities, averaged over (a) days where the over-realisation of risk is in the top $10 \%$ and (b) days where the over-realisation of risk is in the bottom $90 \%$. We compare these cumulative overlaps with a stationary null model where the true covariance matrix is 
$\mathbb{E}_{\text{in}}$.\footnote{A null model with a true covariance matrix equal to the identity $\mathbb{I}$ leads to quite different results, very far from empirical data. This result shows that these overlaps are, unsurprisingly, very sensitive to the underlying correlation structure.} 

The results are quite striking: whereas in most cases, the null model explains rather well the direction in which risk is over-realised, the top $10 \%$ cases are clearly different. For futures, large excess out-of-sample risk is concentrated in the statistical risk modes with the smallest in-sample risk, whereas for stocks, excess risk is in the direction of the statistical modes with the largest in-sample risk (see also the related discussion in \cite{Allez,overlaps}.) 

For futures, risk over-realisation tends to come from the sudden divergence of the spread between tightly correlated contracts, for example associated to the delivery of the same underlying at different expiry dates. The spread between those contracts is typically close to zero, but exogenous shocks might lead such a typically quiet direction to generate anomalous risk along the term structure of the contract. As an example, we observe in Fig. \ref{fig:lambda_vs_t} a spike around April 21st, 2020 that corresponds to the days in which the price of crude oil futures has been strongly stressed by a COVID-induced demand shock. Our metrics thus identifies such directions as fleeting modes, since in the presence of tiny in-sample risk directions, even a moderate out-of-sample risk leads to a very strong spike in the top eigenvalue of $\mathbb{D}$.

In contrast to futures, the absence of strong mechanical correlations between equity instruments leads to a smaller loading of fleeting modes on low risk modes, and a larger loading on high risk modes (industrial sectors and/or equity factors) which tend to over-realize their risk in a systematic fashion. A natural question is  whether known factors could be at the origin of such excess risks  in equity portfolios. A natural candidate is the momentum factor. Indeed, let us consider the case $T_{\text{in}} \gg T_{\text{mom}}$, where $T_{\text{mom}}$ is the time-scale used to build the momentum signal. The {\it in-sample} risk model is then blind to such a factor, because the directions defined by momentum signal randomly rotate over time and average out when $T_{\text{in}} \gg T_{\text{mom}}$. Hence, because of the impact of investors trading in and out \cite{Volpati}, these factors should over-realize their expected risk provided $T_{\text{mom}} \gtrsim T_{\text{out}}$.

In order to quantify the role of factors (including momentum) in the observed excess risk, we define a metric that measures the alignment of a given factor direction with the subspace spanned by the $n$ largest eigenvectors of $\mathbb{D}(t)$. More precisely, we define by the normalized factor loadings $z_{\text{f},i}(t)$ on the real assets $i=1, \ldots, N$ (with $\lVert \vec{z}_\text{f}(t) \rVert^2 = 1$) and similarly rotate the fleeting modes $\vec{z}_a(t)$ back into the real asset basis. We then consider the following overlap 
\be 
\phi_n(t) := \sum_{a=1}^n \left(\vec{z}_\text{f}(t) \cdot
\vec{z}_a(t) \right)^2 \equiv \sum_{a=1}^n \left( \sum_{i=1}^N {z}_{\text{f},i}(t) \, {z}_{a,i}(t) \right)^2.
\ee 
Note that $0 \leq \phi_n(t) \leq 1$, with $\phi_N(t) \equiv 1$, since $\vec{z}_a$ is a complete ortho-normal basis.  

\begin{figure}
    \centering
    \includegraphics{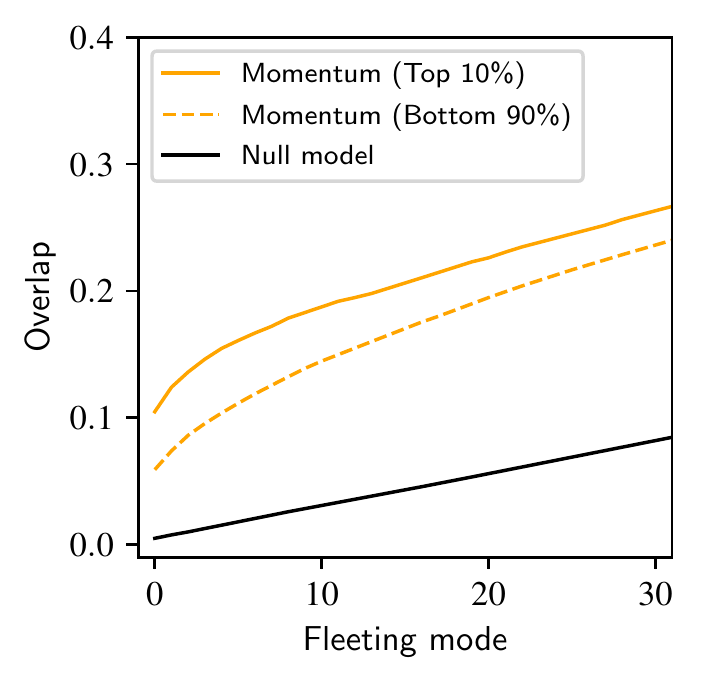}
    \caption{
    Cumulative overlap between the momentum direction and the first 30 fleeting modes in the equity space for the top 10 \% and the bottom 90 \% risk over-realisation. The null model has the same projection amplitude on the statistical risk modes $\vec{v}_\mu(t)$ as momentum, but with randomly scrambled signs. The contribution of the momentum factor to excess risk is clear. 
	}
	   \label{fig:momentum}
\end{figure}

Fig.~\ref{fig:momentum} shows the average value of $\phi_n(t)$ over the whole period, for $n \leq 30$ when the factor $\text{f} = \text{mom}$ is the momentum factor for stocks\footnote{We define momentum as market-neutralized rank-transformed lagged trend signal, with $\textrm{trend}_t = p_t/\textrm{ewma}\left(p_{t},\; \textrm{halflife}=100\textrm{D}\right)$.}. We compare this result with a null model that has the same projection amplitude on the statistical risk modes $\vec{v}_\mu(t)$ as momentum, but with randomly scrambled signs. This graph clearly shows that a significant portion of the risk over-realization in the equity space can indeed be explained as an exposure to the momentum factor, which is itself buffeted by the price impact of momentum trader.


\paragraph{Conclusion} Using Random Matrix Theory, we have provided a universal and versatile tool to analyse the statistical significance and financial origin of risk over-realisation in large portfolios. The eigenvalues and eigenvectors of an appropriately constructed matrix mixing in-sample and out-of-sample data allows one to identify ``fleeting modes'', i.e. portfolios that carry significant excess risk, signalling (ex-post) a change in the correlation structure in the underlying asset space. Our proposed test is furthermore independent of the ``true'' underlying correlation structure, which is obviously unknown to the modeler. We have shown empirically that such fleeting modes exist both in futures markets and in equity markets, and analyzed the directions in which excess risk manifests itself. We have proposed a metric to quantify the alignment between known factors and fleeting modes. As a case in point, momentum exposure clearly appears as a source of excess risk in equity portfolios that is not captured by low frequency correlation matrices.

\end{document}